\documentclass[sn-mathphys]{sn-arXiv}

\jyear{2023}%

\theoremstyle{thmstyleone}%
\theoremstyle{thmstyletwo}%

\theoremstyle{thmstylethree}%

\begin{document}

\title[Non-linear Diffusive Shock Acceleration of Cosmic Rays]{Non-linear Diffusive Shock Acceleration of Cosmic Rays -- Quasi-thermal and Non-thermal Particle Distributions}


\author*[1]{\fnm{Bojan} \sur{Arbutina}}\email{bojan.arbutina@matf.bg.ac.rs}



\affil*[1]{\orgdiv{Department of Astronomy}, \orgname{Faculty of Mathematics, University of Belgrade}, \orgaddress{\street{Studentski trg 16}, \city{Belgrade}, \postcode{11000}, \country{Serbia}}}




\abstract{Diffusive shock acceleration (DSA) of particles at collisionless shocks is {the major} accepted paradigm about the origin of cosmic rays (CRs). As a theory it was developed during the late 1970s in the so-called test-particle case. If one considers the influence of CR particles at shock structure, then we are talking about non-linear DSA. We use semi-analytical Blasi's model of non-linear DSA to obtain non-thermal spectra of both protons and electrons, starting from their quasi-thermal spectra for which we assumed the $\kappa$-distribution, a commonly observed distribution in out-of-equilibrium space plasmas. We treated more carefully than in previous work {the} jump conditions at the subshock {and}  included electron heating, resonant and, additionally, non-resonant magnetic field instabilities produced by CRs in the precursor. Also, corrections for escaping flux of protons and synchrotron losses of electrons have been made.}

\keywords{Acceleration of particles, ISM: cosmic rays, Shock waves, Methods: analytical, Methods: numerical}



\maketitle

\section{Introduction}\label{sec1}

The {main} accepted paradigm that aims to explain the acceleration of particles to cosmic ray (CR) energies up to the $\sim 10^{15}$ eV or even higher  is the so-called diffusive shock acceleration (DSA) at collisionless interstellar medium (ISM) shocks \cite{Longair1994, Morlino2016}. Primary sites of CR acceleration are believed to be supernova remnants (SNRs) and other astrophysical objects in our Galaxy. {As a theory, DSA} was developed during the late 1970s independently by \cite{Axford1977, Krymsky1977, Bell1978a} and \cite{BO1978}. There are two main approaches to the problem, macroscopic and microscopic introduced by Bell \cite{Bell1978a}. In common to both approaches is that they treat the so-called test-particle case, when particles do not affect the shock. If one considers the influence of CR particles at shock structure, then we are talking about non-linear DSA, CR back-reaction and modified shocks (see e.g. \cite{Drury1983, BE1999, MD2001, Blasi2002a, Blasi2002b, AB2005}).

{The} presence of CRs affects the shock in such a way that it changes the (Rankine-Hugoniot) jump conditions, i.e. the very structure of the shock. This can be understood as follows: {high-energy} particles diffuse ahead of the shock and their non-negligible pressure/energy density induce the so-called precursor with density, pressure and velocity gradients. The discontinuity is still present at the so-called subshock with compression $R_\mathrm{sub} = \rho _2/\rho _1 = u _1/u_2$, where $\rho$ is the density and $u$ fluid velocity in the shock frame. However, this compression is smaller than the total compression of a modified shock $R_\mathrm{tot}  = \rho _2/\rho _0 = u _0/u_2$. Indices, 2, 1, 0, mark respectively downstream, immediate upstream and far upstream values with {regard}  to the subshock, with shock velocity in the laboratory frame being $u_s = - u_0$, since far upstream plasma in this frame is assumed to be at rest.

In the test-particle approach the standard DSA particle spectrum is in the power-law form $f(p) \propto p^{-3R/(R-1)} \propto p^{-4}$, for strong shock with compression $R=4$. While CRs modify the shock, {the shock itself at the same time} modifies  {the}  CR particle (power-law) spectrum, producing it to be more concave-up.  {Further upstream the CR particles reach, the} more energy/momentum they have, so the low-energy particles will be confined to the subshock, and high-energy particles can sample the whole precursor. Conditionally speaking, the low-energy particles will experience only the jump at the subshock and have {a} steeper spectrum, while the high-energy particles will experience larger compression, consequently having a flatter spectrum. The overall spectrum will thus be concave-up.

The DSA-based description strictly hold for ions i.e. protons, whose acceleration is more easy to understand since the typical shock thickness should {be} of the order of proton gyro-radius \cite{CS2014a, CS2014b, CS2014c, Zekovic2019}, and thus a fraction of protons needs to be only slightly supra-thermal in order to cross and re-cross the shock unaffected and {engage} in DSA cycles. Because of their much smaller mass, and consequently smaller gyro-radii, {the} acceleration of CR electrons is generally less understood. Nevertheless, kinetic particle-in-cell (PIC) simulations that include both protons and electrons \cite{Pea2015, GG2015, AZ2021a}, as well as synchrotron radio observations (see \cite{Urosevic2014}) suggest that electron spectra resemble those of protons.

In the next section, we shall use semi-analytical Blasi's model of non-linear DSA \cite{Blasi2002a,Blasi2002b} to obtain non-thermal spectra of both protons and electrons, starting from their quasi-thermal spectra for which we assumed the $\kappa$-distribution, a commonly observed distribution in out-of-equilibrium space plasmas \cite{LM2011, Livadiotis2017, Livadiotis2018, LM2022}.

\section{Analysis and Results}\label{sec2}

For modelling proton and electron spectra, we shall use Blasi's semi-analytical model whose details can be found in \cite{Blasi2002a, Blasi2002b} (see also \cite{Blasi2004, AB2005, Bea2005, Bea2007, Ferrand2010, Pavlovic2018, Uea2019, AZ2021a}). Blasi's model implies solving diffusion-advection equation
\begin{equation}
\frac{1}{3}\Big( \frac{1}{R_{\mathrm{tot}}} - U_p\Big) p \frac{d f}{d p} - \Big( U_p + \frac{1}{3} p \frac{d U_p}{d p} \Big) f = 0 ,
\label{eq1}
\end{equation}
coupled with equations of mass and momentum conservation
\begin{equation}
\rho = \frac{\rho _0 u_0}{u} = \rho_0 / U_p,
\end{equation}
\begin{equation}
\rho u^2 + P_\mathrm{CR} + P_\mathrm{w} + P_\mathrm{th}= \rho _0 u_0^2 +P_0,
\label{ec}
\end{equation}
i.e.
\begin{equation}
\Xi _p + \alpha _p+ \Pi _p= 1 + \frac{1}{\gamma M_0^2} - U_p,
\label{ec2}
\end{equation}
where $\rho$ is density,
\begin{equation}
\Pi _p = \frac{P_\mathrm{th}}{P_0} = U_p ^{-\gamma} \Big( 1 + \zeta (\gamma -1) \frac{M_0^2}{M_A} (1-U_p^\gamma)\Big)
\label{ah}
\end{equation}
is normalized thermal gas pressure (see \cite{BE1999}), $\Xi_p$ CR pressure, and $\alpha _p$ magnetic field, i.e. waves pressure, both normalized to shock ram pressure $\rho u_0^2$, and $U_p$ is velocity in the precursor normalized to $u_0$. Mach number and Alfven-Mach number in the unperturbed medium (ISM far upstream) are defined as $M_0 = u_0/c_s$ and $M_A = u_0/v_A$, where $c_s = \sqrt{\gamma P_0/\rho_0}$ is the sound speed, $v_A = B_0/\sqrt{4\pi \rho _0}$ Alfven velocity, $B_0$ ISM magnetic field {strength}, $\gamma$ adiabatic index (set to 5/3) and $\zeta$ is the Alfven-heating parameter \cite{Cea2009}. The model assumes that particles of momentum $p$ typically diffuse up to a distance
\begin{equation}
x_p = \frac{D(p) }{U_p u_0}
\label{D}
\end{equation}
in the precursor, where $D(p)$ is Bohm-like diffusion coefficient, so that all relevant physical quantities $u$, $P_\mathrm{CR}$, $P_\mathrm{w}$, $P_\mathrm{th}$, and consequently $U_p$, $\Xi _p$, $\alpha _p$, $\Pi _p$, depend on $p$.

\subsection{The $\kappa$--distribution}\label{subsec1}

PIC simulations \cite{CS2014a, CS2014b, CS2014c} show that the proton downstream spectrum consists of thermal, supra-thermal and non-thermal parts. Presence of supra-thermal part is expected, since in order to enter DSA cycles, particles need to be pre-accelerated somehow. This can be accomplished by specular reflection, through the so-called shock-drift acceleration (SDA)  by a combination of SDA and DSA \cite{Cea2015, Pea2015}, or by a kind of micro DSA ($\mu$-DSA, \cite{ZA2019}).  As already said, the acceleration of electrons is less easier to understand \cite{MD2001, Artem2023}, but they should generally go through the similar pre-acceleration process and once they reach the injection momentum of protons, they will continue to behave in the same fashion and further accelerate through the DSA mechanism.

In \cite{Cea2015}, in order to explain the downstream particle spectrum, the authors
introduce the so-called minimal model. As assumed in this model, while {the} majority of the (thermal) protons will be advected and isotropized downstream, a constant fraction of them can gain extra energy by performing a few gyrations while drifting along the shock surface, performing SDA cycles. A fraction of these supra-thermal particles provide the seed (injection) particles for the standard DSA mechanism. The model thus describes
 supra-thermal and non-thermal particle distributions through the same formalism, as basically the same distribution.

 One could also try to describe thermal and supra-thermal particle distribution with one continuous quasi-thermal distribution -- the $\kappa$-distribution \cite{AZ2021b, Arbutina2023}
 \begin{equation}
\frac{dN}{dp} =  4 \pi p^2 f(p) = \frac{N_0 4 \pi p^2}{(\pi \kappa p_\kappa ^2)^{3/2}} \frac{\Gamma(\kappa +1)}{\Gamma(\kappa -\frac{1}{2})} \frac{1}{\Big[ 1+ \frac{p^2}{\kappa p_{\kappa} ^2} \Big]^{\kappa +1}},\ \ \ p_\kappa ^2 = 2 mkT_\kappa
\label{k1}
\end{equation}
  In this quasi-thermal distribution, index $\kappa$ is a free parameter which serves as a kind of a measure of non-equilibrium \cite{LM2011, Livadiotis2017}. When $\kappa \to \infty$, the plasma reaches equilibrium and the distribution becomes Maxwellian
  \begin{equation}
\frac{dN}{dp} =  \frac{4\pi p^2 N_0}{(2\pi m kT)^{3/2}} e^{-\frac{p^2}{2mkT}}.
\end{equation}
  As inferred from PIC simulations, this happens further from the shock, in the far downstream \cite{AZ2021b} and { possibly} immediately behind the shock after enough time {has passed}. For higher momenta, $\kappa$-distribution is actually a power-law with index $-2\kappa$. Note that while $T$ is true thermodynamic temperature, $T_\kappa$ in Eq.(\ref{k1}) is not.

At some injection momentum, $\kappa$-distribution should match non-thermal distribution.
We can thus find the matching condition that relates injection parameter $\xi ={p_\mathrm{inj}}/{p_\mathrm{th}}$ ($p_\kappa ^2 = \frac{\kappa -3/2}{\kappa} p_{\mathrm{th}}^2$) and
injection efficiency $\eta=n_\mathrm{CR}/n$, where $n$ is the total particle number density and $n_\mathrm{CR}$ that of CRs \cite{Arbutina2023}
\begin{equation}
{\eta} = \frac{4}{3\sqrt{\pi}} (R_{\mathrm{\mathrm{sub}}}-1) \frac{\Gamma(\kappa +1)}{(\kappa - \frac{3}{2})^{3/2}\Gamma(\kappa -\frac{1}{2})} \frac{\xi ^3}{\Big[ 1+ \frac{\xi^2}{\kappa -\frac{3}{2}} \Big]^{\kappa +1}} .
\label{eta1}
\end{equation}
When $\kappa \to \infty$ one obtains standard injection efficiency (from matching non-thermal to thermal (Maxwell) distribution) as \cite{Bea2005}
\begin{equation}
\eta = \frac{4}{3\sqrt{\pi}} (R_{\mathrm{\mathrm{sub}}}-1) \xi ^3 e^{-\xi ^2}.
\label{eta2}
\end{equation}

\subsection{Magnetic fields from streaming instability}\label{subsec2}

\subsubsection{Resonant instabilities}\label{subsubsec1}

In addition to CR acceleration at strong collisionless ISM shock, a process that is {happening in parellel} is the magnetic field amplification. Some amplification must occur, since  plain shock compression (of normal field component $B_\perp$) cannot explain the observed magnetic field strengths in synchrotron sources, for example.
Magnetic field pressure in Eq. (\ref{ec}) emerges from the so-called streaming instability induced by CRs, that can be resonant or non-resonant \cite{Bell1978a, Bell2004, AB2009}.

In the case of resonant instability, the unstable modes are Alfven waves whose wavelength is assumed to {be} in resonance with CR gyration or Larmor radius $r_L = p_\perp /(e B_\parallel)$, where $e$ is elementary charge and $B_\parallel \approx B_0$, i.e. for the wave number $k$ we should have $kr_L \sim 1$. The stationary equation for the growth and transport of self-generated Alfven waves \cite{McKV1982} with normalized pressure $\alpha _r = \frac{B^2}{8 \pi \rho _0 u_0^2}$ can be transformed to \cite{Cea2009}
\begin{equation}
2 U_p \frac{d \alpha _r}{dx} = (1-\zeta) V_A \frac{d\Xi}{dx} - 3 \alpha _r \frac{dU_p}{dx} ,
\end{equation}
where $V_A$ is normalized compressed Alfven speed $V_A = \sqrt{U_p}/M_A$. Assuming strong shocks i.e. high Mach number, with dominant CR pressure upstream, $\Xi \approx 1 - U_p$, the last equation can be solved to give \cite{Cea2009}
\begin{equation}
\alpha _r = (1-\zeta) U_p^{-3/2} \frac{1-U_p^2}{4 M_A}.
\label{res}
\end{equation}
The dependence $U_p ^{-3/2}$ describes adiabatic compression, while the term $(1-\zeta)$  regulates Alfven waves damping, {as the corresponding} parameter $0 \le \zeta \le 1$  in  Eq. (\ref{ah}) regulates the amount of Alfven heating. It is clear that this term can not be too small for magnetic field to be substantially amplified.

At the subshock, taking into account {transmission} and reflection of waves, the jump condition for the magnetic field pressure is \cite{VS1999, Ferrand2010, Cea2009}
\begin{equation}
\alpha _{r,2} =   R_\mathrm{sub}^2 \alpha _{r,1}.
\end{equation}

\subsubsection{Non-resonant instabilities}\label{subsubsec2}

Non-resonant or Bell's instabilities \cite{Bell2004} represent almost purely growing modes
that do not correspond to Alfven waves. Bell estimated the saturated field to be \cite{Bell2004}
 \begin{equation}
\frac{B_\mathrm{sat}^2}{8\pi} \approx \frac{1}{2} \frac{u_0}{c} u_\mathrm{CR},
 \label{sat}
 \end{equation}
where $u_\mathrm{CR} = \frac{1}{\gamma _\mathrm{CR}-1} P_\mathrm{CR}$ is CR energy density. In ref. \cite{Sea2017} the authors added this non-resonant term to resonantly amplified magnetic field $\frac{B_\mathrm{res}^2}{8\pi} \approx u_\mathrm{CR} /M_A $ to describe magnetic {field} amplification, particle acceleration and synchrotron emission of SNRs, but only in the test-particle regime of DSA, without paying attention to the actual structure of the precursor. {Empirical evidence that the theory presented by \cite{Sea2017} is quite inaccurate in reproducing radio fluxes of supernova remnants was given by \cite{Leahy2022}. }
In \cite{Pea2018}, full hydrodynamic modeling was performed and synchrotron radio evolution of SNRs was investigated, with non-linear particle acceleration based on Blasi's model, but aside from different global dependence ($\propto u_0/c$ instead $1/M_A$) the non-resonant instabilities were treated in a similar fashion to resonant (Eq. (\ref{res})) which can not be {correct}.

A rigorous description of these
 instabilities is still missing, but we can use the arguments by \cite{Bell2004} and assume {for the normalized magnetic pressure due to non-resonant instabilities $\alpha _n = \frac{B_\mathrm{res}^2}{8\pi \rho_0 u_0^2} = \frac{3 \iota}{2} \frac{u_0}{c} \Xi,$
} where $\iota$ is some parameter of order unity.
If again one assumes $\Xi \approx 1 - U_p$, we have finally
 \begin{equation}
\alpha _n = \frac{3 \iota}{2} \frac{u_0}{c} (1- U_p).
\label{non}
 \end{equation}
 The lack of understanding of the interaction of particles with these instabilities, {does not permit} us to go much further than this in their quantitative description. For the same reasons, it is difficult to treat magnetic field jump conditions at the subshock, and we can only assume the normal field component(s) to be compressed, so that
 \begin{displaymath}
\alpha _{n,2} =  \frac{1 + (j-1) R_\mathrm{sub}^2}{j} \alpha _{n,1}, \ \ \ j = 1+ \frac{B_{1, \perp }^2}{ B_{1, \parallel}^2} =\left\{ \begin{array}{ll}
2 & \textrm{\ if\ $B_{1, \parallel} = B_{1, \perp} $}\\
3 & \textrm{\ if\  $B_{1, x} = B_{1, y} = B_{1, z}$}
\end{array} \right.
\end{displaymath}
In \cite{Sea2017} the authors assumed random (isotropic) distribution upstream with $j=3$, but it may as well be that $j=2$, or something else.

It is worth noting that $\alpha _n \gg \alpha _r$ only for large shock velocities, so the non-resonant instabilities should be relevant e.g. in the early stages of evolution of SNRs \cite{AB2009}. However, this will depend on the exact values of different parameters involved. By using Eqs. (\ref{res}) and (\ref{non}), for $\iota = 1$ and $\zeta = 0$, $\alpha _r= \alpha _n$ gives
\begin{equation}
\frac{6u_0^2}{v_A c} = U_p^{-3/2} + U_p^{-1/2}.
 \end{equation}
The transition between dominant non-resonant and resonant instabilities would then be at $u_0 \sim 1000-2000$ km/s for $v_A \sim 10$ km/s. Nevertheless, in this intermediate domain both instabilities should be relevant, as it can be seen in Fig. 1. Panel (a) shows the test particle case with shock velocity $u_0$ = 1010 km/s and $1/R_\mathrm{prec} = 0.95$, i.e. $R_\mathrm{tot} \gtrsim R_\mathrm{sub} = R \approx 4$, and  panel (b) represents a modified shock with $u_0$ = 1150 km/s and $1/R_\mathrm{prec} = 0.3$.  Since $U_p$ can be related to $u(x)$ i.e. through Eq. (\ref{D}) to the position in the precursor $x$, $U_1 = 1/R_\mathrm{prec} = R_\mathrm{sub} / R_\mathrm{tot}$ marks the position of the subshock, while unperturbed medium starts i.e. precursor ends at $U = U_0 = 1$. We can see that in the intermediate, transition domain, resonant instabilities dominate ahead of the subshock, and non-resonant further upstream.

\begin{figure}[h]
\center{
  \includegraphics[bb=0 0 771 641,angle=0,width=0.85\textwidth,keepaspectratio]{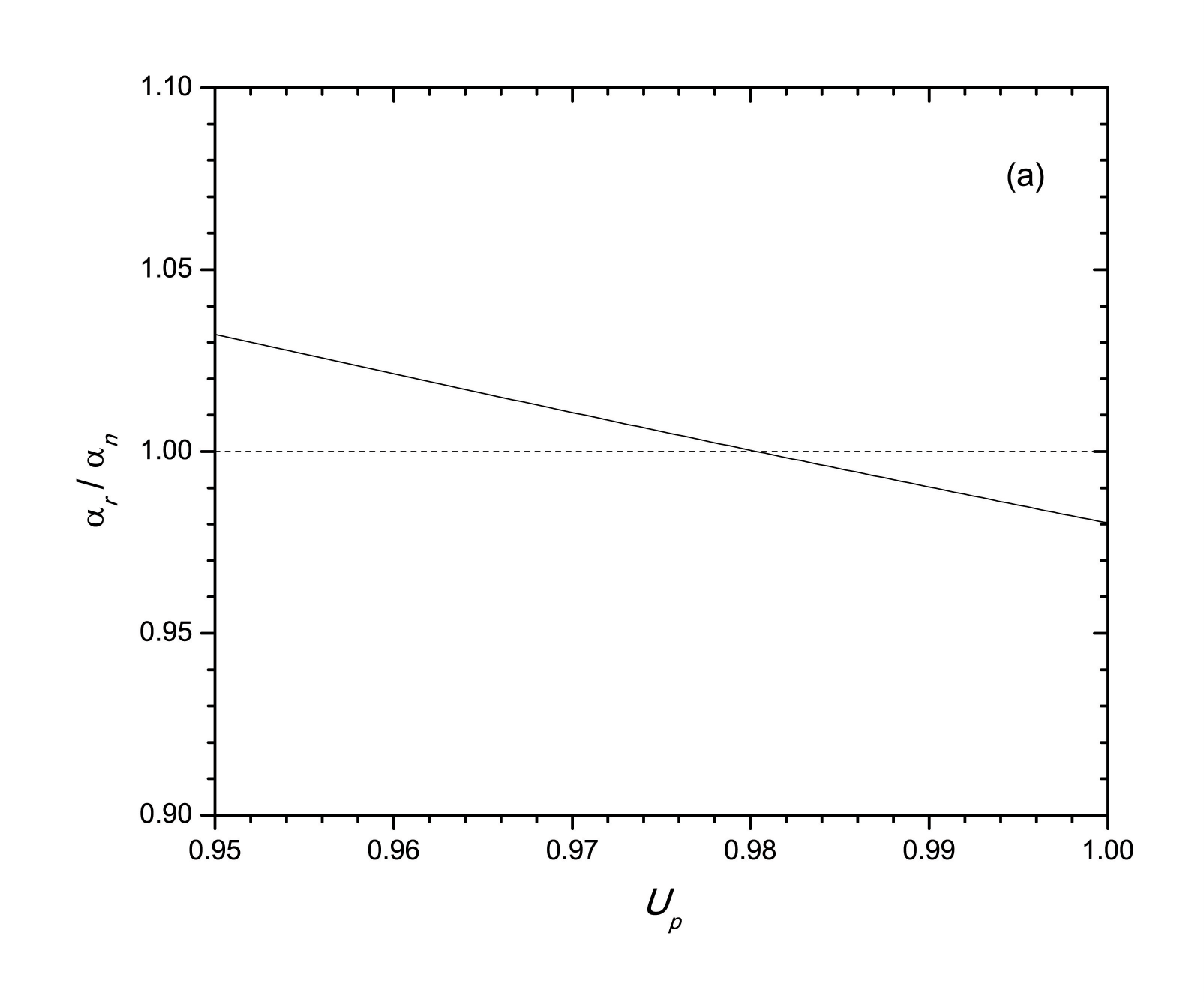}
  \includegraphics[bb=0 0 771 641,angle=0,width=0.85\textwidth,keepaspectratio]{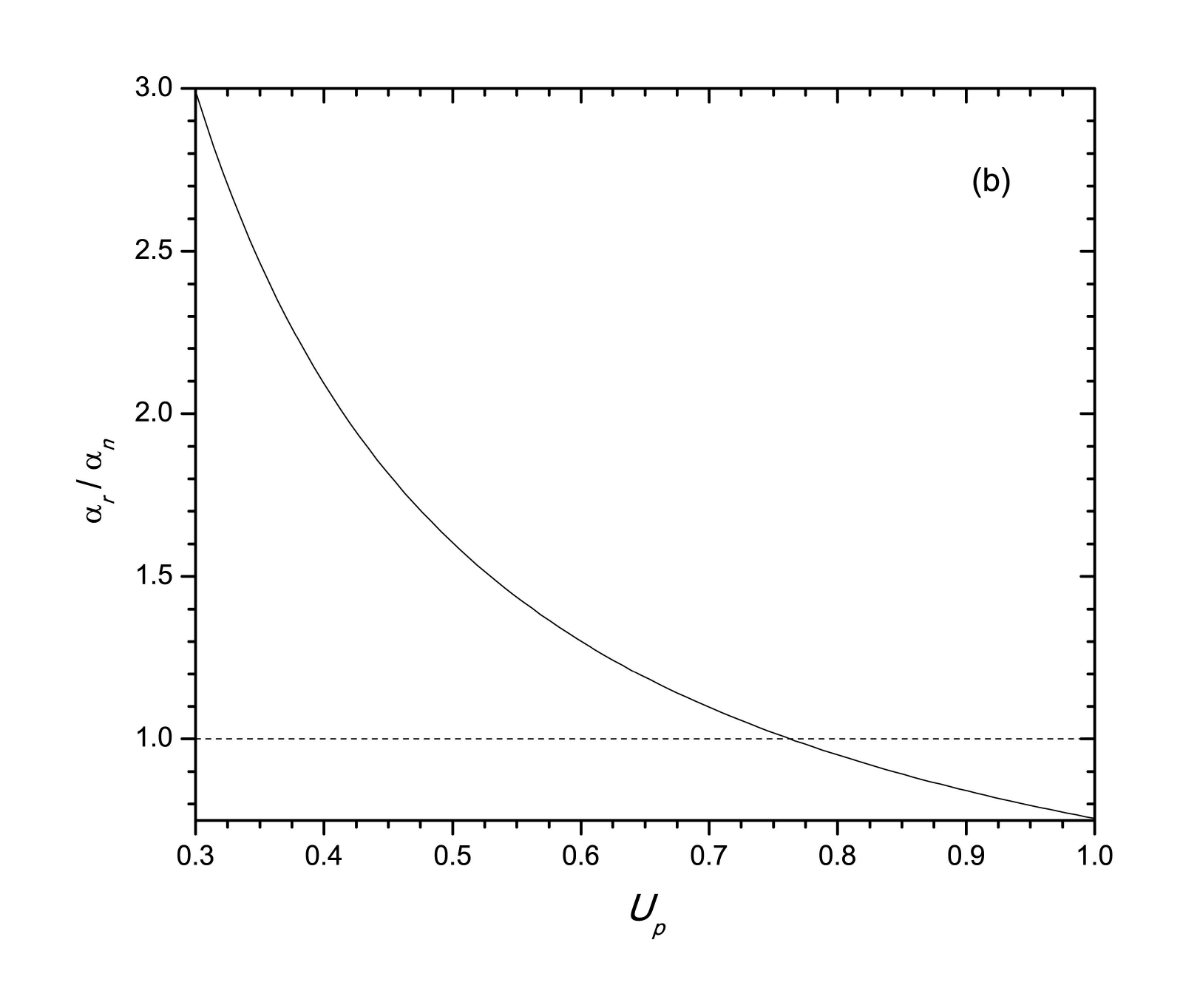}
  \caption{{Ratio of magnetic field pressure from resonant instabilites to that from non-resonant instabilities (solid line).} The dashed line determines a value of $U_p$ i.e. the position in the precursor where $\alpha _r= \alpha _n$.  Top panel (a) shows the test particle case with shock velocity $u_0$ = 1010 km/s and $1/R_\mathrm{prec} = 0.95$, while the bottom panel (b) represents a modified shock with $u_0$ = 1150 km/s and $1/R_\mathrm{prec} = 0.3$.  For both cases, we have assumed $v_A \sim 10$ km/s, $\iota = 1$ and $\zeta = 0$. }}
\end{figure}

\subsection{Electron heating}\label{subsec3}

In order to find injection momentum of particles entering DSA through Eq. (\ref{eta1}) or (\ref{eta2}) one needs to know downstream temperature. We shall assume $\eta _e = \eta _p$, i.e. $\xi _e = \xi _p$ ($\kappa _e = \kappa _p$), however $T_{2,e} \neq T_{2,p}$. Treating electrons independently (more precisely, energy conservation for electrons), from Rankine-Hugoniot jump conditions one expects $\beta = T_{2,e} / T_{2,p} = m_e/m_p$, nevertheless observations of SNRs show that this is not the case and that there is significant electron heating, expected to {be} happening in the precursor \cite{Gh2007, Gh2013, Gh2016}. {Recently, apparent discrepancies in $\beta$ between results from SNR shocks, Solar wind shocks and PIC simulations were highlighted in \cite{Ray2023}.}

For intermediate {SNR} shock velocities, {the} amount of heating {appears to be} roughly constant $\Delta E \approx 0.3$ keV, so that \cite{Gh2016}
\begin{equation}
\beta \approx \frac{\frac{3}{16} m_e u_0^2 + \Delta E}{\frac{3}{16} m_p u_0^2}.
\label{beta1}
\end{equation}

In refs. \cite{AZ2021a, Arbutina2023} the authors implemented this by removing the energy  $\Delta E = 0.3$ keV from Alfven-heated protons and {adding it} to electrons (constant electron heating ahead of the subshock), so that the downstream temperatures are $T_{2,p} = T'_{2,p} -  \Delta E /k$, $T_{2,e} = T'_{2,e} -  \Delta E /k$ (where temperatures $T'_2$ are obtained from jump conditions). Nevertheless, it seems that there is flattening in $\beta = \beta (u_0)$ dependence, so that for high shock velocities {$\beta \to \beta _0 \approx $ const} \cite{Gh2016}. This means that for very strong shocks the energy $\Delta E$ may {be}  a constant fraction of shock ram pressure or proton downstream temperature rather {the} constant itself.

We shall try to account for this by assuming that
\begin{equation}
\Delta E = a kT_{2,p} + b.
\label{dE}
\end{equation}
Downstream temperatures for strong shocks can be found from energy conservation, and if we add/subtract $\Delta E$, we have
\begin{equation}
kT_{2,e} = \frac{\gamma -1}{2\gamma} \Big( 1 - \frac{1}{R^2} \Big) m_e u_0^2 + a kT_{2,p} + b,
\end{equation}
\begin{equation}
kT_{2,p} = \frac{\gamma -1}{2\gamma} \Big( 1 - \frac{1}{R^2} \Big) m_p u_0^2 - a kT_{2,p} - b,
\end{equation}
{yielding the} ratio
\begin{equation}
\beta = \frac{ \frac{\gamma -1}{2\gamma} \Big( 1 - \frac{1}{R^2} \Big) \Big((1+a)m_e  +a m_p\Big) u_0^2 + b}{\frac{\gamma -1}{2\gamma} \Big( 1 - \frac{1}{R^2} \Big) m_p u_0^2  - b}.
\label{beta2}
\end{equation}
From the last equation we see that
\begin{equation}
\Delta E _0 = b = 0.3\ \mathrm{keV},
\end{equation}
and in accordance with observations \cite{Gh2016}
\begin{equation}
\beta _0 = a + (1+a)m_e/m_p \approx a = 0.05.
\end{equation}
Functional dependence in Eq. (\ref{beta2}), as well as that in Eq. (\ref{beta1}) with $\Delta E$ = const, are plotted together in Fig 2.

\begin{figure}[h]
\center{
  \includegraphics[bb=0 0 771 641,angle=0,width=0.99\textwidth,keepaspectratio]{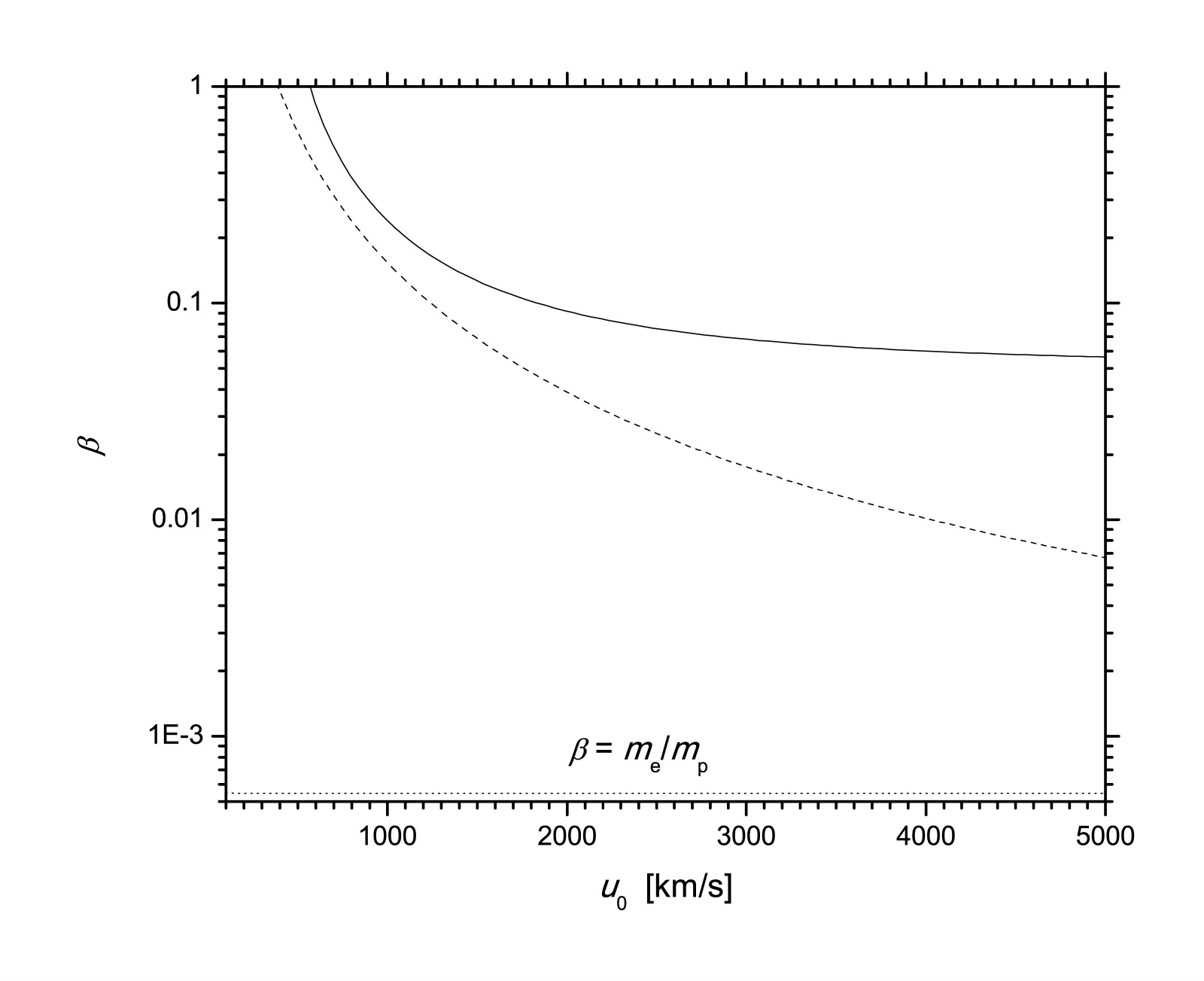}
  \caption{Ratio of downstream temperatures of electrons and protons as a function of shock velocity. Solid line represents functional dependence in Eq. (\ref{beta2}), dashed line that in Eq. (\ref{beta1}) with $\Delta E$ = const, while the dotted line represent the expected value from jump conditions, assuming no heating i.e. energy conservation for electrons. }}
\end{figure}

In our modeling, Eq. (\ref{dE}) i.e. $\Delta E = \beta _0 kT_{2,p} + \Delta E_0$ can be applied at the subshock, so that
\begin{equation}
T_{2,e} = \frac{\gamma -1}{2k\gamma} \Big( \frac{1}{R_\mathrm{prec}^2} - \frac{1}{R_\mathrm{tot}^2} \Big) m_e u_0^2 +  T_{1,e} + \beta _0 T_{2,p} + \Delta E_0 /k.
\end{equation}
On the other hand, from the overall jump conditions, including Alfven waves \cite{VS1999, Ferrand2010, Pavlovic2018}, and Bell instabilities one has
\begin{eqnarray}
\frac{T_{2,p} + T_{2,e}}{T_{1,p} + T_{1,e}}  &=&  \frac{(\gamma+1)R_{\rm sub}-(\gamma-1)\left[1-(R_\mathrm{sub}-1)\Delta \right]}{\left[(\gamma+1) - (\gamma-1)R_{\rm sub}\right]R_{\rm sub}} = A,\\
\Delta &=& (R_{\rm sub}-1)^2\frac{P_{r,1} + \frac{j-1}{j} P_{n,1}}{P_{p,1}+P_{e,1}} =(R_{\rm sub}-1)^2 \Delta ^*.\nonumber
\end{eqnarray}
The last two equations with \cite{AZ2021a}
\begin{equation}
\frac{T_{1,p}}{T_0}  = R_\mathrm{prec}^{\gamma -1} \Big( 1 + \zeta (\gamma -1) \frac{M_0^2}{M_A} (1-R_\mathrm{prec}^{-\gamma})\Big), \ \ \
\frac{T_{1,e}}{T_0}  = R_\mathrm{prec}^{\gamma -1},
\end{equation}
assuming plain adiabatic compression for electrons, allow us to calculate proton and electron downstream temperatures:
\begin{eqnarray}
\frac{T_{2,p}}{T_{0}}  &=&  \frac{1}{1+\beta _0} \Big( \frac{T_{1,p}}{T_0} + \frac{T_{1,e}}{T_0}  \Big)  A -  \frac{1}{1+\beta _0} B
,\\
\frac{T_{2,e}}{T_{0}}  &=&  \frac{\beta_0}{1+\beta _0} \Big( \frac{T_{1,p}}{T_0} + \frac{T_{1,e}}{T_0}  \Big) A +  \frac{1}{1+\beta _0} B, \nonumber
\end{eqnarray}
where
\begin{equation}
B = \frac{\gamma -1}{2} \Big( \frac{1}{R_\mathrm{prec}^2} - \frac{1}{R_\mathrm{tot}^2} \Big) \frac{m_e}{m_p} M_0^2 +  \frac{T_{1,e}}{T_0}  + \frac{\Delta E_0}{kT_0}.
\end{equation}

\subsection{Modeling}\label{subsec4}

In order to obtain particle spectra, we need to solve simultaneously Eq. (\ref{eq1}) for protons and electrons, and {the} differentiated momentum equation (Eq. (\ref{ec2}))
\begin{equation}
\frac{1}{3}\Big( \frac{1}{R_{\mathrm{tot}}} - U_p\Big) p \frac{{d} f_{p}}{{d} p} - \Big( U_p + \frac{1}{3} p \frac{{d} U_p}{{d} p} \Big) f_p= 0 ,
\end{equation}
\begin{equation}
\frac{1}{3}\Big( \frac{1}{R_{\mathrm{tot}}} - U_p\Big) p \frac{{d} f_{e}}{{d} p} - \Big( U_p + \frac{1}{3} p \frac{{d} U_p}{{d} p} \Big) f_e= 0 ,
\end{equation}
\begin{eqnarray}
\frac{{d}U_p}{{d}p}\left[1 - \frac{U_p^{-(\gamma+1)}}{ M_{0}^2}\left(2 + \zeta(\gamma -1)\frac{M_{0}^2}{M_{A}}\right) - \frac{1-\zeta}{8 M_{A}} \frac{U_p^2+3}{U_p^{5/2}} - \frac{3 \iota}{2} \frac{u_0}{c} \right]\nonumber \\
=  \frac{p^4 f_p}{\sqrt{1 + p^2}} + \frac{p^4 f_e}{\sqrt{(m_{e}/m_{p})^2  + p^2 }}.
\label{syst}
\end{eqnarray}
In the last set of equations we introduced dimensionless quantities $\frac{p}{m_{p} c} \rightarrow p$, $\frac{4\pi}{3} \frac{m_{p}^4 c^5}{\rho _0 u_0^2} f \rightarrow f$, { included} both resonant and non-resonant instabilities, and {used} Eq. (\ref{ah}) for the pressure of thermal protons, while for thermal electrons, as already said, we assumed adiabatic compression $\Pi _e = U_p ^{-\gamma} $ (with $\Delta E$ added at the subshock). Note that $M_{0}$ and $M_{A}$ are (proton) Mach number and Alfven-Mach number in the far upstream.

Between $p _{{\rm inj},e}$ and $p _{{\rm inj},p}$ it is assumed that $U_p = \frac{1}{R_{\mathrm{prec}}}$, so that $f_e \propto p^{-3R_{\rm{sub}}/(R_{\rm{sub}}-1)}$.
We need to prescribe $R_{\rm prec}$ here and at the {beginning}  of integration ($p = p _{{\rm inj},p}$,
$f_p =  \frac{3 R_{\rm{tot}}}{R_{\rm{sub}} - 1} \frac{\eta n_{0}}{4\pi p_{{\rm inj},p}^3}$, $U_p = 1/R_{\rm{prec}}$) and start iterations. In each iteration $R_{\mathrm{sub}}$ (and consequently $R_{\mathrm{tot}}$) is calculated through equations \citep{Blasi2002a, Blasi2002b, Ferrand2010}.
\begin{eqnarray}
R_{\mathrm{sub}} ^2 & - & \frac{2+(\gamma -1 + 2\gamma P_{w1}^*) M_1^2}{2(\gamma -2)P_{w1}^* M_1^2 } R_{\mathrm{sub}} + \frac{\gamma +1}{2(\gamma -2)P_{w1}^* } =0, \nonumber \\
M_1^2 &=&  M_0^2 R_{\mathrm{prec}}^{-\gamma -1} \left(2 + \zeta(\gamma -1)\frac{M_{0}^2}{M_{A}} (1-  R_{\mathrm{prec}}^{-\gamma})\right)^{-1}  , \\
P_{w1}^*&=& \frac{1}{\gamma M_1^2} {\Delta ^*}= (R_{\mathrm{prec}} -1) \left(\frac{1-\zeta}{4 M_{A}} \sqrt{R_{\mathrm{prec}}} (R_{\mathrm{prec}}+1) + \frac{j-1}{j} \frac{3\iota}{2} \frac{u_0}{c}\right)
\nonumber .
\end{eqnarray}
Iterative procedure is stopped when $U_p =1$ at
\begin{equation} p_{{\rm max}}
 = \frac{3}{8} \frac{u_0}{c^2} e B_0 R,
\end{equation}
where $R$ is now (SNR) shock radius (see \citep{Bell2013} and
references therein). We actually need to integrate only advection-diffusion equation for protons since $f_e = K_{ep} f_p$, where electron-to-proton ratio at high energies is
$K_{ep} = \frac{\eta _e}{\eta _p} (\frac{m_e}{m_p} \beta)^\frac{3}{2(R_{\rm{sub}}-1) }$ \citep{AZ2021a}.

{We shall apply} two corrections a posteriori: for protons we shall apply a correction due to the flux of escaping particles, and for electrons a {cut-off} at $p_{{\rm loss}}$ due to the synchrotron losses.

\subsubsection{Escaping flux}\label{subsubsec3}

The proton spectrum obtained through the above integration has a sharp break at $p=p_{{\rm max}} $. In reality, we expect particles {to} escape freely after reaching the outer boundary of precursor at $x_{\rm max}$. This can be accounted for by inclusion of an additional {term} $\phi_{\rm esc}$ in Eq. (\ref{eq1}) \citep{Caprioli2010, Ferrand2010},  which will lead to a relatively gradual decrease around $p_{\rm max}$ in the spectrum.

If one defines normalized escape flux $\Phi _{\rm esc} =\frac{\phi_{\rm esc}}{u_0 f}$, the solution of Eq. (\ref{eq1}) can be written as \citep{Caprioli2010, Ferrand2010}
\begin{equation}
f = \frac{3}{U_p - 1/R_{\rm tot}} \frac{\eta n_0}{4 \pi p_{\rm inj}^3} \exp \Bigg( -\int _{p_{\rm inj}} ^p \frac{3(U_p + \Phi _{\rm esc} )}{U_p -1/R_{\rm tot}} \frac{dp'}{p'}\Bigg).
\end{equation}

By using Eq. (\ref{D}) and assuming Bohm diffusion with coefficient $D=\frac{1}{3} \frac{pv}{e B_0}$, we approximate the normalized escape flux as
\begin{equation}
\Phi _{\rm esc} = \frac{1}{{\rm e}^{ \frac{p_{\rm max} c}{p v}} -1},
\end{equation}
from which it follows that the spectrum obtained through the system of equations (\ref{syst}) $f_{p,0}$ can be corrected approximately as
\begin{equation}
f_{p} = f_{p,0}  \big(1 -  {\rm e}^{-\frac{p_{\rm max}}{p}} \big)^{\frac{3R_{\rm tot}}{R_{\rm tot} -1}\frac{p}{p_{\rm max}}},
\end{equation}
where for $p > p_{\rm max}$ we assumed $f_{p,0} \propto p^{\frac{-3R_{\rm tot}}{R_{\rm tot}-1}}$.

\subsubsection{Synchrotron losses}\label{subsubsec4}

Ultra-relativistic electrons in strong magnetic field will emit synchrotron radiation. We should thereby also try to correct electron spectrum for synchrotron losses. Since the magnetic field is strongest at the subshock (particularly downstream), we can reasonably expect that the losses will be dominant there.
We shall therefore use test-particle approach results by \citep{Blasi2010} and \citep{ZA2007} who included an additional {term} in the advection-diffusion equation:
$\frac{1}{p^2}\frac{\partial}{\partial p} (p^2 A f)$. For high-energy electrons suffering bremsstrahlung losses $A\propto p$, while for synchrotron or inverse-Compton losses $A\propto p^2$ \cite{Longair1994}.

In \citep{ZA2007}, by assuming Bohm difussion and taking $A =\frac{4 e^4 B^2}{9 m_e^4 c^6} p^2$, the authors find that the resulting spectrum at high momenta has the form
$f \propto  \sqrt{p}\cdot {\rm e}^{-\big(\frac{p}{p_{\rm loss}}\big)^2}$,
where, after our small adaptation,
\begin{equation}
p_{\rm loss} = \frac{\sqrt{R_B}}{1+ \sqrt{R_B}} \frac{R_{\rm sub} -1}{R_{\rm tot}} \frac{m_e^2 c^2 u_0}{\sqrt{2 e^3 B_2/27}}
\end{equation}
and $R_B = B_2/B_1$ is the magnetic field jump at the subshock.

If we {adapt, highly provisionally,} the procedure previously applied to protons, we can postulate
\begin{equation}
\Phi _{\rm loss} = - \frac{\frac{7 R_{\rm tot} -R_{\rm loss}}{6 R_{\rm tot}R_{\rm loss}} (\frac{p}{p_s})^s}{(\frac{p}{p_s})^s +1} + \frac{2(R_{\rm tot} - R_{\rm loss})}{3 R_{\rm tot} R_{\rm loss}} \Big( \frac{p}{p_{\rm loss}}\Big)^2,
\end{equation}
where compression $R_{\rm loss}$ corresponds to momentum $p_{\rm loss}$. This gives
\begin{equation}
f_{e} = f_{e,0}  \Big[ 1 +  \Big( \frac{p}{p_s}\Big)^{s}  \Big]^
{\frac{7 R_{\rm tot} -R_{\rm loss}}{2s (R_{\rm tot} -R_{\rm loss})}}  {\rm e}^{-\big(\frac{p}{p_{\rm loss}}\big)^2} .
\end{equation}
This agrees with analytical approximation given by \citep{ZA2007} for the test particle case, where $R_{\rm tot} = R_{\rm sub} \approx 4$, $R_{\rm loss} =1$, {which}  have correct limits
\begin{displaymath}
f \propto  \left\{ \begin{array}{ll}
p^{-4} & \textrm{\ if\ $p_{{\rm inj}} \leftarrow p$,}\\
 {p}^{1/2} {\rm e}^{-\big(\frac{p}{p_{\rm loss}}\big)^2} & \textrm{\ if\  $p \to p_{{\rm loss}}$.}
\end{array} \right.
\end{displaymath}

Index $s$ and momentum $p_s$, depend on $R_B$ and $R_{\rm tot}$ ($R_{\rm sub}$) \citep{ZA2007}, however, there is no general analytical expression for them.
Furthermore, for specific $p_s$, particularly if $R_{\rm tot} > 4$, there is a visible pile-up/bump in the spectrum in the cut-off region (see case $R_{\rm tot}=7$ in \citep{Blasi2010}).
Nevertheless, since for strong unmodified shocks we do not expect pronounced bumps, and for modified shock {the} spectrum is already concave-up, we shall only incorporate exponential cut-off at $p_{{\rm loss}}$, so that
\begin{equation}
f_{e} \approx f_{e,0}  {\rm e}^{-\big(\frac{p}{p_{\rm loss}}\big)^2} ,
\end{equation}
where for $p > p_{\rm loss}$ we assumed $f_{e,0}  = K_{ep} f_{p,0}$.

\subsubsection{Results}\label{subsubsec5}

The setup is similar as in \cite{Caprioli2010, AZ2021a, Arbutina2023} and for resonant modes {it} assumes Alfven-heating parameter $\zeta$ = 0.5 {(between extremes $\zeta =0$ -- no heating, and $\zeta =1$ -- maximum heating i.e. complete waves damping)}, for non-resonant modes $\iota =1$, $j=3$, shock velocity $u_0$ = 5000 km/s, ambient density $n_\mathrm{0} \sim$ 0.1 cm$^{-3}$, temperature $T_0 = 10^5$ K, magnetic field $B_0$ = 5.3775 $\mu$Ga, Mach and Alfven-Mach numbers $M_0 = M_A = 135$. We assumed that {the}  injection parameter $\xi$ (and thereby efficiency $\eta$), as well as the index $\kappa$, is the same for protons and electrons.

In Figs. 3 and 4 we give the results for two cases: ($\kappa \to \infty$, $\xi$ = 3.3), and ($\kappa =5$, $\xi$ = 5). Both cases show strongly modified shock/non-linear DSA spectra. The former case gives the proton and electron spectra that match Maxwellians at $p_{\rm inj}$, while the latter case gives spectra that match $\kappa$-distributions. In \cite{Arbutina2023} we already noted that Eq. (\ref{eta1}) generally gives higher efficiency when compared to the case $\kappa \to \infty$ for the same $\xi$ (although this parameter for the $\kappa$-distribution is not uniquely defined, see \cite{Arbutina2023}). This means that in contrast to the Maxwellian-match situations where realistic $\xi \sim  3.5-4$ \cite{Bea2005, CS2014a}, to reach the test particle case for $\kappa$--distribution-match, $\xi$  must be much larger \cite{Arbutina2023}. {Efficiency} $\eta$ in the latter case depends on both $\xi$ and $\kappa$.

 For the case ($\kappa \to \infty$, $\xi$ = 3.3), the subshock compression is $R_\mathrm{sub}$ = 3.17, the total compressions is $R_\mathrm{tot}$ = 7.07, and consequently $R_\mathrm{prec} = 2.23$.  Injection efficiency is $\eta$ = 0.0011 and electron-to-proton ratio at high energies is $K_\mathrm{ep}$ = 0.0011. For the case ($\kappa =5$, $\xi$ = 5),  the subshock compression, the total compression and the precursor compression are, respectively, $R_\mathrm{sub}$ = 3.01, $R_\mathrm{tot}$ = 7.48, $R_\mathrm{prec} = 2.49$, injection efficiency is $\eta$ = 0.0010 and $K_\mathrm{ep}$ = 0.0007.

 {A} more careful treatment of subshock jump conditions and electron heating leads to higher electron-to-proton ratio $K_\mathrm{ep}$ than in our previous work \cite{AZ2021a}. This means that the observed ratio for primary (Galactic) CRs $K_\mathrm{ep} \sim 0.01$ is more easily achievable in these models. The $K_\mathrm{ep}$ ratio is slightly lower for the  $\kappa$--distribution case, for the practically same injection efficiency. In \cite{AZ2021b} we showed that while the distribution behind the shock can be represented by the quasi-thermal $\kappa$--distribution, farther downstream index $\kappa$ increases, the distribution at high momenta becomes steeper, and the (thermal) spectrum tends to Maxwellian. This may also happen with time, during the course of evolution of SNRs, so that the $\kappa$--distribution may be relevant only in the early stages when we still have a non-equilibrium plasma.

 Also, in contrast to previous calculations that considered only resonant instabilities \cite{AZ2021a, Arbutina2023}, we included Bell's instabilities that actually prevail at $u_s$ = 5000 km/s. Since the CR energy density in the precursor dominates over magnetic field and thermal gas energies, this does not affect {the shape of the spectrum and overall parameters} as much as the subshock jump condition (with electron heating), but may be important in practical applications of CR astrophysics, dealing with magnetic field, e.g. in modelling of gamma or synchrotron radio emission of astrophysical sources.

\begin{figure}[h]
\center{
  \includegraphics[bb=0 0 771 641,angle=0,width=0.99\textwidth,keepaspectratio]{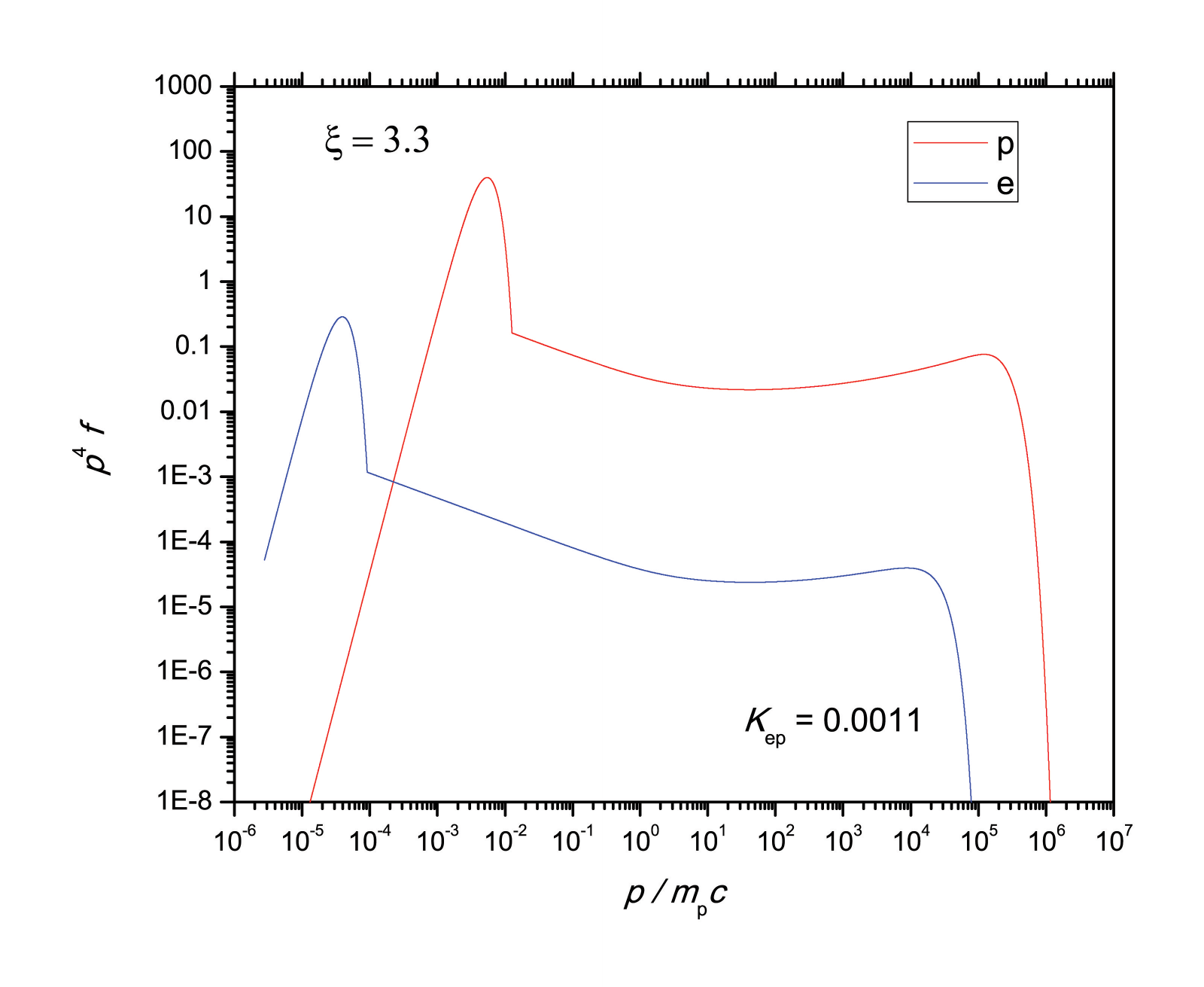}
  \caption{Non-thermal proton and electron spectra for injection parameter $\xi$ = 3.3 that match thermal Maxwell distribution. }}
\end{figure}

\begin{figure}[h]
\center{
  \includegraphics[bb=0 0 771 641,angle=0,width=0.99\textwidth,keepaspectratio]{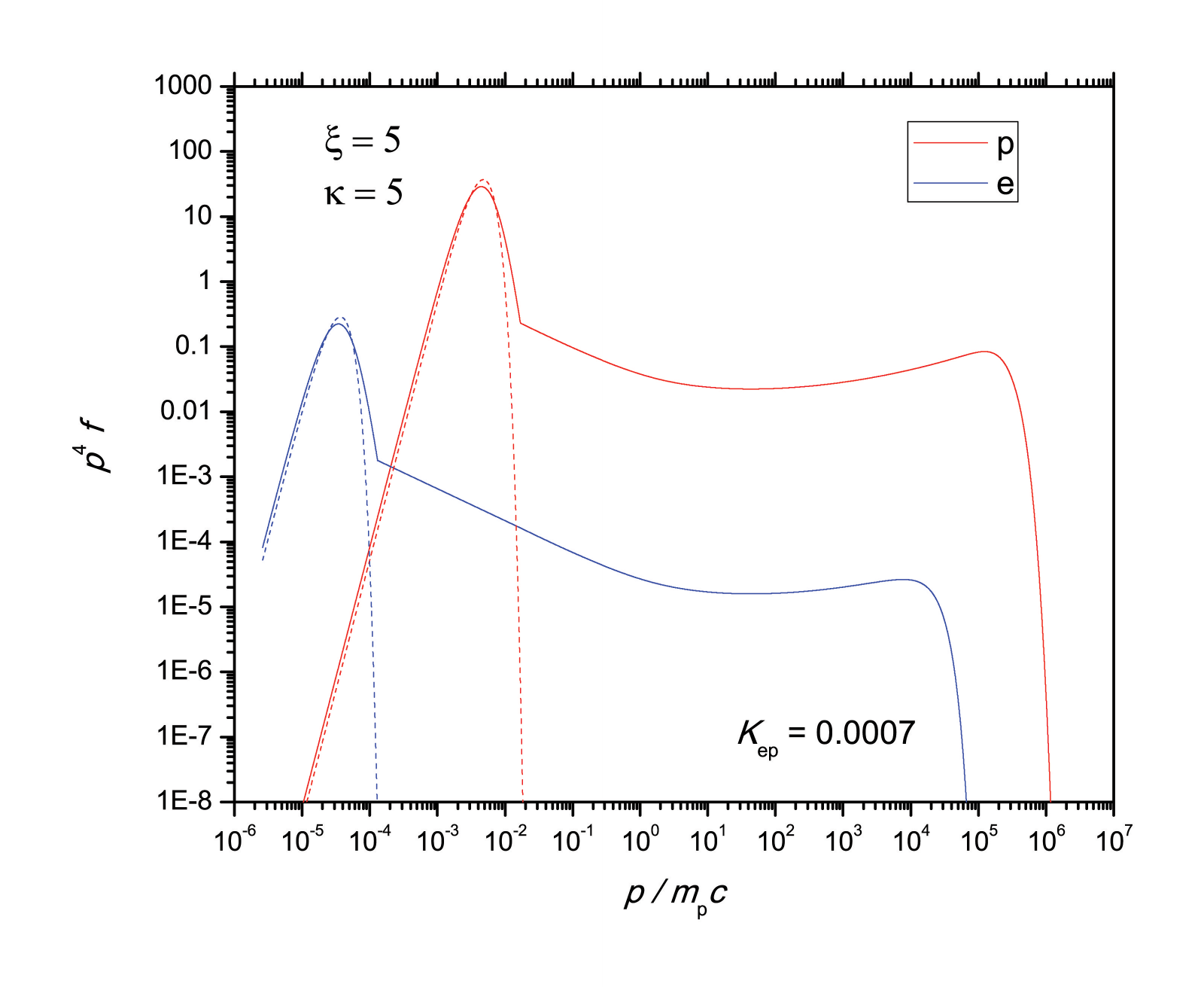}
  \caption{Non-thermal proton and electron spectra for injection parameter $\xi$ = 5 that match quasi-thermal $\kappa$--distribution with index $\kappa = 5$. Solid lines show $\kappa$-distributions and non-thermal distributions that join at injection momenta $p_\mathrm{inj}$. Maxwellians with the same downstream temperatures ($T_p$ and $T_e$) are shown with dashed lines. }}
\end{figure}

\section{Conclusion}\label{sec13}

In this paper we gave an overview of non-linear DSA, based on Blasi's semi-analytical model, while incorporating some add-ons, in addition to corrections for escaping flux of protons and synchrotron losses of electrons, that should be important for obtaining more realistic protons and electrons non-thermal spectra. We started by assuming a quasi-thermal $\kappa$-distribution at low energies \cite{AZ2021b}, that through matching condition at the injection momentum provides different recipe for calculating injection efficiency $\eta$. This recipe generally gives higher efficiency when compared to the Maxwellian for the same injection parameter $\xi$ \cite{Arbutina2023}. We treated more carefully than in previous work {the} jump conditions at the subshock, included electron heating, resonant and, additionally, non-resonant magnetic field instabilities \cite{Bell2004}.

Since {these} instabilities should be dominant for shock velocities $u_s > 1000-2000$ km/s, their inclusion, as well as correct estimate of electron-to-proton ratio at high energies $K_\mathrm{ep}$, are extremely important for e.g. the evolution of young SNRs and their synchrotron radio emission modeling (see e.g. \cite{Pea2018}). To this {end}, it is crucial to address the questions how injection efficiencies of both protons and electrons, downstream electron-to-proton temperature ratio, and consequently CR electron-to-proton ratio at high energies, change with shock velocity. Along with better theoretical understanding of injection and acceleration processes, some answers may hopefully be provided by PIC simulations.


\section*{Acknowledgments}

The author acknowledges the funding provided through the contract No. 451-03-47/2023-01/200104 by the Ministry of Science, Technological Development and Innovation of the Republic of Serbia, and through the joint project of the Serbian Academy of Sciences and Arts and Bulgarian Academy of Sciences on the detection of Galactic and extragalactic SNRs and HII regions.

\bibliography{EPJ-arXiv}


\begin{thebibliography}{51}
\ifx \bisbn   \undefined \def \bisbn  #1{ISBN #1}\fi
\ifx \binits  \undefined \def \binits#1{#1}\fi
\ifx \bauthor  \undefined \def \bauthor#1{#1}\fi
\ifx \batitle  \undefined \def \batitle#1{#1}\fi
\ifx \bjtitle  \undefined \def \bjtitle#1{#1}\fi
\ifx \bvolume  \undefined \def \bvolume#1{\textbf{#1}}\fi
\ifx \byear  \undefined \def \byear#1{#1}\fi
\ifx \bissue  \undefined \def \bissue#1{#1}\fi
\ifx \bfpage  \undefined \def \bfpage#1{#1}\fi
\ifx \blpage  \undefined \def \blpage #1{#1}\fi
\ifx \burl  \undefined \def \burl#1{\textsf{#1}}\fi
\ifx \doiurl  \undefined \def \doiurl#1{\url{https://doi.org/#1}}\fi
\ifx \betal  \undefined \def \betal{\textit{et al.}}\fi
\ifx \binstitute  \undefined \def \binstitute#1{#1}\fi
\ifx \binstitutionaled  \undefined \def \binstitutionaled#1{#1}\fi
\ifx \bctitle  \undefined \def \bctitle#1{#1}\fi
\ifx \beditor  \undefined \def \beditor#1{#1}\fi
\ifx \bpublisher  \undefined \def \bpublisher#1{#1}\fi
\ifx \bbtitle  \undefined \def \bbtitle#1{#1}\fi
\ifx \bedition  \undefined \def \bedition#1{#1}\fi
\ifx \bseriesno  \undefined \def \bseriesno#1{#1}\fi
\ifx \blocation  \undefined \def \blocation#1{#1}\fi
\ifx \bsertitle  \undefined \def \bsertitle#1{#1}\fi
\ifx \bsnm \undefined \def \bsnm#1{#1}\fi
\ifx \bsuffix \undefined \def \bsuffix#1{#1}\fi
\ifx \bparticle \undefined \def \bparticle#1{#1}\fi
\ifx \barticle \undefined \def \barticle#1{#1}\fi
\bibcommenthead
\ifx \bconfdate \undefined \def \bconfdate #1{#1}\fi
\ifx \botherref \undefined \def \botherref #1{#1}\fi
\ifx \url \undefined \def \url#1{\textsf{#1}}\fi
\ifx \bchapter \undefined \def \bchapter#1{#1}\fi
\ifx \bbook \undefined \def \bbook#1{#1}\fi
\ifx \bcomment \undefined \def \bcomment#1{#1}\fi
\ifx \oauthor \undefined \def \oauthor#1{#1}\fi
\ifx \citeauthoryear \undefined \def \citeauthoryear#1{#1}\fi
\ifx \endbibitem  \undefined \def \endbibitem {}\fi
\ifx \bconflocation  \undefined \def \bconflocation#1{#1}\fi
\ifx \arxivurl  \undefined \def \arxivurl#1{\textsf{#1}}\fi
\csname PreBibitemsHook\endcsname

\bibitem{Longair1994}
\begin{bbook}
\bauthor{\bsnm{{Longair}}, \binits{M.S.}}:
\bbtitle{High Energy Astrophysics. Vol.2: Stars, the Galaxy and the
  Interstellar Medium}
vol. \bseriesno{2},
(\byear{1994})
\end{bbook}
\endbibitem

\bibitem{Morlino2016}
\begin{bchapter}
\bauthor{\bsnm{Morlino}, \binits{G.}}:
\bctitle{High-energy cosmic rays from supernovae}.
In: \beditor{\bsnm{Alsabti}, \binits{A.W.}},
\beditor{\bsnm{Murdin}, \binits{P.}} (eds.)
\bbtitle{Handbook of Supernovae},
pp. \bfpage{1711}--\blpage{1736}.
\bpublisher{Springer},
\blocation{Cham}
(\byear{2016})
\end{bchapter}
\endbibitem

\bibitem{Axford1977}
\begin{bchapter}
\bauthor{\bsnm{{Axford}}, \binits{W.I.}},
\bauthor{\bsnm{{Leer}}, \binits{E.}},
\bauthor{\bsnm{{Skadron}}, \binits{G.}}:
\bctitle{{The Acceleration of Cosmic Rays by Shock Waves}}.
In: \bbtitle{International Cosmic Ray Conference}.
\bsertitle{International Cosmic Ray Conference},
vol. \bseriesno{11},
p. \bfpage{132}
(\byear{1977})
\end{bchapter}
\endbibitem

\bibitem{Krymsky1977}
\begin{barticle}
\bauthor{\bsnm{{Krymskii}}, \binits{G.F.}}:
\batitle{{A regular mechanism for the acceleration of charged particles on the
  front of a shock wave}}.
\bjtitle{Akademiia Nauk SSSR Doklady}
\bvolume{234},
\bfpage{1306}--\blpage{1308}
(\byear{1977})
\end{barticle}
\endbibitem

\bibitem{Bell1978a}
\begin{barticle}
\bauthor{\bsnm{{Bell}}, \binits{A.R.}}:
\batitle{{The acceleration of cosmic rays in shock fronts - I.}}
\bjtitle{Mon. Not. R. Astron. Soc.}
\bvolume{182},
\bfpage{147}--\blpage{156}
(\byear{1978}).
\doiurl{10.1093/Mon. Not. R. Astron. Soc./182.2.147}
\end{barticle}
\endbibitem

\bibitem{BO1978}
\begin{barticle}
\bauthor{\bsnm{{Blandford}}, \binits{R.D.}},
\bauthor{\bsnm{{Ostriker}}, \binits{J.P.}}:
\batitle{{Particle acceleration by astrophysical shocks.}}
\bjtitle{Astrophys. J. Lett.}
\bvolume{221},
\bfpage{29}--\blpage{32}
(\byear{1978}).
\doiurl{10.1086/182658}
\end{barticle}
\endbibitem

\bibitem{Drury1983}
\begin{barticle}
\bauthor{\bsnm{{Drury}}, \binits{L.O.}}:
\batitle{{An introduction to the theory of diffusive shock acceleration of
  energetic particles in tenuous plasmas}}.
\bjtitle{Reports on Progress in Physics}
\bvolume{46}(\bissue{8}),
\bfpage{973}--\blpage{1027}
(\byear{1983}).
\doiurl{10.1088/0034-4885/46/8/002}
\end{barticle}
\endbibitem

\bibitem{BE1999}
\begin{barticle}
\bauthor{\bsnm{{Berezhko}}, \binits{E.G.}},
\bauthor{\bsnm{{Ellison}}, \binits{D.C.}}:
\batitle{{A Simple Model of Nonlinear Diffusive Shock Acceleration}}.
\bjtitle{Astrophys. J.}
\bvolume{526}(\bissue{1}),
\bfpage{385}--\blpage{399}
(\byear{1999}).
\doiurl{10.1086/307993}
\end{barticle}
\endbibitem

\bibitem{MD2001}
\begin{barticle}
\bauthor{\bsnm{{Malkov}}, \binits{M.A.}},
\bauthor{\bsnm{{Drury}}, \binits{L.O.}}:
\batitle{{Nonlinear theory of diffusive acceleration of particles by shock
  waves}}.
\bjtitle{Reports on Progress in Physics}
\bvolume{64}(\bissue{4}),
\bfpage{429}--\blpage{481}
(\byear{2001}).
\doiurl{10.1088/0034-4885/64/4/201}
\end{barticle}
\endbibitem

\bibitem{Blasi2002a}
\begin{barticle}
\bauthor{\bsnm{{Blasi}}, \binits{P.}}:
\batitle{{A novel approach to non linear Shock acceleration}}.
\bjtitle{Nuclear Physics B Proceedings Supplements}
\bvolume{110},
\bfpage{475}--\blpage{477}
(\byear{2002})
{\href{https://arxiv.org/abs/astro-ph/0111529}{{arXiv:astro-ph/0111529}}}
{[astro-ph]}.
\doiurl{10.1016/S0920-5632(02)01539-6}
\end{barticle}
\endbibitem

\bibitem{Blasi2002b}
\begin{barticle}
\bauthor{\bsnm{{Blasi}}, \binits{P.}}:
\batitle{{A semi-analytical approach to non-linear shock acceleration}}.
\bjtitle{Astroparticle Physics}
\bvolume{16}(\bissue{4}),
\bfpage{429}--\blpage{439}
(\byear{2002})
{\href{https://arxiv.org/abs/astro-ph/0104064}{{arXiv:astro-ph/0104064}}}
{[astro-ph]}.
\doiurl{10.1016/S0927-6505(01)00127-X}
\end{barticle}
\endbibitem

\bibitem{AB2005}
\begin{barticle}
\bauthor{\bsnm{{Amato}}, \binits{E.}},
\bauthor{\bsnm{{Blasi}}, \binits{P.}}:
\batitle{{A general solution to non-linear particle acceleration at
  non-relativistic shock waves}}.
\bjtitle{Mon. Not. R. Astron. Soc.}
\bvolume{364}(\bissue{1}),
\bfpage{76}--\blpage{80}
(\byear{2005})
{\href{https://arxiv.org/abs/astro-ph/0509673}{{arXiv:astro-ph/0509673}}}
{[astro-ph]}.
\doiurl{10.1111/j.1745-3933.2005.00110.x}
\end{barticle}
\endbibitem

\bibitem{CS2014a}
\begin{barticle}
\bauthor{\bsnm{{Caprioli}}, \binits{D.}},
\bauthor{\bsnm{{Spitkovsky}}, \binits{A.}}:
\batitle{{Simulations of Ion Acceleration at Non-relativistic Shocks. I.
  Acceleration Efficiency}}.
\bjtitle{Astrophys. J.}
\bvolume{783}(\bissue{2}),
\bfpage{91}
(\byear{2014})
{\href{https://arxiv.org/abs/1310.2943}{{arXiv:1310.2943}}}
{[astro-ph.HE]}.
\doiurl{10.1088/0004-637X/783/2/91}
\end{barticle}
\endbibitem

\bibitem{CS2014b}
\begin{barticle}
\bauthor{\bsnm{{Caprioli}}, \binits{D.}},
\bauthor{\bsnm{{Spitkovsky}}, \binits{A.}}:
\batitle{{Simulations of Ion Acceleration at Non-relativistic Shocks. II.
  Magnetic Field Amplification}}.
\bjtitle{Astrophys. J.}
\bvolume{794}(\bissue{1}),
\bfpage{46}
(\byear{2014})
{\href{https://arxiv.org/abs/1401.7679}{{arXiv:1401.7679}}}
{[astro-ph.HE]}.
\doiurl{10.1088/0004-637X/794/1/46}
\end{barticle}
\endbibitem

\bibitem{CS2014c}
\begin{barticle}
\bauthor{\bsnm{{Caprioli}}, \binits{D.}},
\bauthor{\bsnm{{Spitkovsky}}, \binits{A.}}:
\batitle{{Simulations of Ion Acceleration at Non-relativistic Shocks. III.
  Particle Diffusion}}.
\bjtitle{Astrophys. J.}
\bvolume{794}(\bissue{1}),
\bfpage{47}
(\byear{2014})
{\href{https://arxiv.org/abs/1407.2261}{{arXiv:1407.2261}}}
{[astro-ph.HE]}.
\doiurl{10.1088/0004-637X/794/1/47}
\end{barticle}
\endbibitem

\bibitem{Zekovic2019}
\begin{barticle}
\bauthor{\bsnm{{Zekovi{\'c}}}, \binits{V.}}:
\batitle{{Resonant micro-instabilities at quasi-parallel collisionless shocks:
  Cause or consequence of shock (re)formation}}.
\bjtitle{Physics of Plasmas}
\bvolume{26}(\bissue{3}),
\bfpage{032106}
(\byear{2019})
{\href{https://arxiv.org/abs/1903.01169}{{arXiv:1903.01169}}}
{[astro-ph.HE]}.
\doiurl{10.1063/1.5050909}
\end{barticle}
\endbibitem

\bibitem{Pea2015}
\begin{barticle}
\bauthor{\bsnm{{Park}}, \binits{J.}},
\bauthor{\bsnm{{Caprioli}}, \binits{D.}},
\bauthor{\bsnm{{Spitkovsky}}, \binits{A.}}:
\batitle{{Simultaneous Acceleration of Protons and Electrons at Nonrelativistic
  Quasiparallel Collisionless Shocks}}.
\bjtitle{Phys. Rev. Lett.}
\bvolume{114}(\bissue{8}),
\bfpage{085003}
(\byear{2015})
{\href{https://arxiv.org/abs/1412.0672}{{arXiv:1412.0672}}}
{[astro-ph.HE]}.
\doiurl{10.1103/PhysRevLett.114.085003}
\end{barticle}
\endbibitem

\bibitem{GG2015}
\begin{barticle}
\bauthor{\bsnm{{Guo}}, \binits{F.}},
\bauthor{\bsnm{{Giacalone}}, \binits{J.}}:
\batitle{{The Acceleration of Electrons at Collisionless Shocks Moving Through
  a Turbulent Magnetic Field}}.
\bjtitle{Astrophys. J.}
\bvolume{802}(\bissue{2}),
\bfpage{97}
(\byear{2015})
{\href{https://arxiv.org/abs/1409.5854}{{arXiv:1409.5854}}}
{[astro-ph.HE]}.
\doiurl{10.1088/0004-637X/802/2/97}
\end{barticle}
\endbibitem

\bibitem{AZ2021a}
\begin{barticle}
\bauthor{\bsnm{{Arbutina}}, \binits{B.}},
\bauthor{\bsnm{{Zekovi{\'c}}}, \binits{V.}}:
\batitle{{Non-linear diffusive shock acceleration: A recipe for injection of
  electrons}}.
\bjtitle{Astroparticle Physics}
\bvolume{127},
\bfpage{102546}
(\byear{2021})
{\href{https://arxiv.org/abs/2012.15117}{{arXiv:2012.15117}}}
{[astro-ph.HE]}.
\doiurl{10.1016/j.astropartphys.2020.102546}
\end{barticle}
\endbibitem

\bibitem{Urosevic2014}
\begin{barticle}
\bauthor{\bsnm{{Uro{\v{s}}evi{\'c}}}, \binits{D.}}:
\batitle{{On the radio spectra of supernova remnants}}.
\bjtitle{Astrophys. Space Sci.}
\bvolume{354}(\bissue{2}),
\bfpage{541}--\blpage{552}
(\byear{2014})
{\href{https://arxiv.org/abs/1408.1107}{{arXiv:1408.1107}}}
{[astro-ph.HE]}.
\doiurl{10.1007/s10509-014-2095-4}
\end{barticle}
\endbibitem

\bibitem{LM2011}
\begin{barticle}
\bauthor{\bsnm{{Livadiotis}}, \binits{G.}},
\bauthor{\bsnm{{McComas}}, \binits{D.J.}}:
\batitle{{Invariant Kappa Distribution in Space Plasmas Out of Equilibrium}}.
\bjtitle{Astrophys. J.}
\bvolume{741}(\bissue{2}),
\bfpage{88}
(\byear{2011}).
\doiurl{10.1088/0004-637X/741/2/88}
\end{barticle}
\endbibitem

\bibitem{Livadiotis2017}
\begin{bchapter}
\bauthor{\bsnm{{Livadiotis}}, \binits{G.}}:
\bctitle{{Statistical origin and properties of kappa distributions}}.
In: \bbtitle{Journal of Physics Conference Series}.
\bsertitle{Journal of Physics Conference Series},
vol. \bseriesno{900},
p. \bfpage{012014}
(\byear{2017}).
\doiurl{10.1088/1742-6596/900/1/012014}
\end{bchapter}
\endbibitem

\bibitem{Livadiotis2018}
\begin{barticle}
\bauthor{\bsnm{{Livadiotis}}, \binits{G.}}:
\batitle{{Kappa Distributions: Statistical Physics and Thermodynamics of Space
  and Astrophysical Plasmas}}.
\bjtitle{Universe}
\bvolume{4}(\bissue{12}),
\bfpage{144}
(\byear{2018}).
\doiurl{10.3390/universe4120144}
\end{barticle}
\endbibitem

\bibitem{LM2022}
\begin{barticle}
\bauthor{\bsnm{{Livadiotis}}, \binits{G.}},
\bauthor{\bsnm{{McComas}}, \binits{D.J.}}:
\batitle{{Physical Correlations Lead to Kappa Distributions}}.
\bjtitle{Astrophys. J.}
\bvolume{940}(\bissue{1}),
\bfpage{83}
(\byear{2022})
{\href{https://arxiv.org/abs/2210.05752}{{arXiv:2210.05752}}}
{[physics.plasm-ph]}.
\doiurl{10.3847/1538-4357/ac99df}
\end{barticle}
\endbibitem

\bibitem{Blasi2004}
\begin{barticle}
\bauthor{\bsnm{{Blasi}}, \binits{P.}}:
\batitle{{Nonlinear shock acceleration in the presence of seed particles}}.
\bjtitle{Astroparticle Physics}
\bvolume{21}(\bissue{1}),
\bfpage{45}--\blpage{57}
(\byear{2004})
{\href{https://arxiv.org/abs/astro-ph/0310507}{{arXiv:astro-ph/0310507}}}
{[astro-ph]}.
\doiurl{10.1016/j.astropartphys.2003.10.008}
\end{barticle}
\endbibitem

\bibitem{Bea2005}
\begin{barticle}
\bauthor{\bsnm{{Blasi}}, \binits{P.}},
\bauthor{\bsnm{{Gabici}}, \binits{S.}},
\bauthor{\bsnm{{Vannoni}}, \binits{G.}}:
\batitle{{On the role of injection in kinetic approaches to non-linear particle
  acceleration at non-relativistic shock waves}}.
\bjtitle{Mon. Not. R. Astron. Soc.}
\bvolume{361}(\bissue{3}),
\bfpage{907}--\blpage{918}
(\byear{2005})
{\href{https://arxiv.org/abs/astro-ph/0505351}{{arXiv:astro-ph/0505351}}}
{[astro-ph]}.
\doiurl{10.1111/j.1365-2966.2005.09227.x}
\end{barticle}
\endbibitem

\bibitem{Bea2007}
\begin{barticle}
\bauthor{\bsnm{{Blasi}}, \binits{P.}},
\bauthor{\bsnm{{Amato}}, \binits{E.}},
\bauthor{\bsnm{{Caprioli}}, \binits{D.}}:
\batitle{{The maximum momentum of particles accelerated at cosmic ray modified
  shocks}}.
\bjtitle{Mon. Not. R. Astron. Soc.}
\bvolume{375}(\bissue{4}),
\bfpage{1471}--\blpage{1478}
(\byear{2007})
{\href{https://arxiv.org/abs/astro-ph/0612424}{{arXiv:astro-ph/0612424}}}
{[astro-ph]}.
\doiurl{10.1111/j.1365-2966.2006.11412.x}
\end{barticle}
\endbibitem

\bibitem{Ferrand2010}
\begin{botherref}
\oauthor{\bsnm{Ferrand}, \binits{G.}}:
Blasi's semi-analytical kinetic model of non-linear diffusive shock
  acceleration.
Personal notes
(2010)
\end{botherref}
\endbibitem

\bibitem{Pavlovic2018}
\begin{botherref}
\oauthor{\bsnm{Pavlovi\'c}, \binits{M.Z.}}:
Modeling the radio-evolution of supernova remnants by using hydrodynamic
  simulations and non-linear diffusive shock acceleration.
PhD thesis, University of Belgrade
(2018)
\end{botherref}
\endbibitem

\bibitem{Uea2019}
\begin{barticle}
\bauthor{\bsnm{{Uro{\v{s}}evi{\'c}}}, \binits{D.}},
\bauthor{\bsnm{{Arbutina}}, \binits{B.}},
\bauthor{\bsnm{{Oni{\'c}}}, \binits{D.}}:
\batitle{{Particle acceleration in interstellar shocks}}.
\bjtitle{Astrophys. Space Sci.}
\bvolume{364}(\bissue{10}),
\bfpage{185}
(\byear{2019})
{\href{https://arxiv.org/abs/1910.06006}{{arXiv:1910.06006}}}
{[astro-ph.HE]}.
\doiurl{10.1007/s10509-019-3669-y}
\end{barticle}
\endbibitem

\bibitem{Cea2009}
\begin{barticle}
\bauthor{\bsnm{{Caprioli}}, \binits{D.}},
\bauthor{\bsnm{{Blasi}}, \binits{P.}},
\bauthor{\bsnm{{Amato}}, \binits{E.}},
\bauthor{\bsnm{{Vietri}}, \binits{M.}}:
\batitle{{Dynamical feedback of self-generated magnetic fields in cosmic ray
  modified shocks}}.
\bjtitle{Mon. Not. R. Astron. Soc.}
\bvolume{395}(\bissue{2}),
\bfpage{895}--\blpage{906}
(\byear{2009})
{\href{https://arxiv.org/abs/0807.4261}{{arXiv:0807.4261}}}
{[astro-ph]}.
\doiurl{10.1111/j.1365-2966.2009.14570.x}
\end{barticle}
\endbibitem

\bibitem{Cea2015}
\begin{barticle}
\bauthor{\bsnm{{Caprioli}}, \binits{D.}},
\bauthor{\bsnm{{Pop}}, \binits{A.-R.}},
\bauthor{\bsnm{{Spitkovsky}}, \binits{A.}}:
\batitle{{Simulations and Theory of Ion Injection at Non-relativistic
  Collisionless Shocks}}.
\bjtitle{Astrophys. J. Lett.}
\bvolume{798}(\bissue{2}),
\bfpage{28}
(\byear{2015})
{\href{https://arxiv.org/abs/1409.8291}{{arXiv:1409.8291}}}
{[astro-ph.HE]}.
\doiurl{10.1088/2041-8205/798/2/L28}
\end{barticle}
\endbibitem

\bibitem{ZA2019}
\begin{bchapter}
\bauthor{\bsnm{{Zekovic}}, \binits{V.}},
\bauthor{\bsnm{{Arbutina}}, \binits{B.}}:
\bctitle{{Quasi-parallel collisionless shock (re)formation and particle
  acceleration by (non)resonant micro-instabilities}}.
In: \bbtitle{Supernova Remnants: An Odyssey in Space After Stellar Death II},
p. \bfpage{129}
(\byear{2019})
\end{bchapter}
\endbibitem

\bibitem{Artem2023}
\begin{barticle}
\bauthor{\bsnm{{Bohdan}}, \binits{A.}}:
\batitle{{Electron acceleration in supernova remnants}}.
\bjtitle{Plasma Physics and Controlled Fusion}
\bvolume{65}(\bissue{1}),
\bfpage{014002}
(\byear{2023})
{\href{https://arxiv.org/abs/2211.13992}{{arXiv:2211.13992}}}
{[astro-ph.HE]}.
\doiurl{10.1088/1361-6587/aca5b2}
\end{barticle}
\endbibitem

\bibitem{AZ2021b}
\begin{barticle}
\bauthor{\bsnm{{Arbutina}}, \binits{B.}},
\bauthor{\bsnm{{Zekovi{\'c}}}, \binits{V.}}:
\batitle{{On the distribution function of suprathermal particles at
  collisionless shocks}}.
\bjtitle{Journal of High Energy Astrophysics}
\bvolume{32},
\bfpage{65}--\blpage{70}
(\byear{2021})
{\href{https://arxiv.org/abs/2108.09085}{{arXiv:2108.09085}}}
{[physics.plasm-ph]}.
\doiurl{10.1016/j.jheap.2021.08.003}
\end{barticle}
\endbibitem

\bibitem{Arbutina2023}
\begin{bchapter}
\bauthor{\bsnm{{Arbutina}}, \binits{B.}}:
\bctitle{{Diffusive Shock Acceleration of Cosmic Rays -- Quasi-thermal and
  Non-thermal Particle Distributions}}.
In: \bbtitle{11th International Conference of the Balkan Physical Union
  (BPU11), 28 August - 1 September 2022, Belgrade, Serbia}
(\byear{2023})
\end{bchapter}
\endbibitem

\bibitem{Bell2004}
\begin{barticle}
\bauthor{\bsnm{{Bell}}, \binits{A.R.}}:
\batitle{{Turbulent amplification of magnetic field and diffusive shock
  acceleration of cosmic rays}}.
\bjtitle{Mon. Not. R. Astron. Soc.}
\bvolume{353}(\bissue{2}),
\bfpage{550}--\blpage{558}
(\byear{2004}).
\doiurl{10.1111/j.1365-2966.2004.08097.x}
\end{barticle}
\endbibitem

\bibitem{AB2009}
\begin{barticle}
\bauthor{\bsnm{{Amato}}, \binits{E.}},
\bauthor{\bsnm{{Blasi}}, \binits{P.}}:
\batitle{{A kinetic approach to cosmic-ray-induced streaming instability at
  supernova shocks}}.
\bjtitle{Mon. Not. R. Astron. Soc.}
\bvolume{392}(\bissue{4}),
\bfpage{1591}--\blpage{1600}
(\byear{2009})
{\href{https://arxiv.org/abs/0806.1223}{{arXiv:0806.1223}}}
{[astro-ph]}.
\doiurl{10.1111/j.1365-2966.2008.14200.x}
\end{barticle}
\endbibitem

\bibitem{McKV1982}
\begin{barticle}
\bauthor{\bsnm{{McKenzie}}, \binits{J.F.}},
\bauthor{\bsnm{{V\"olk}}, \binits{H.J.}}:
\batitle{{Non-linear theory of cosmic ray shocks including self-generated
  Alfven waves}}.
\bjtitle{Astron. Astrophys.}
\bvolume{116}(\bissue{2}),
\bfpage{191}--\blpage{200}
(\byear{1982})
\end{barticle}
\endbibitem

\bibitem{VS1999}
\begin{barticle}
\bauthor{\bsnm{{Vainio}}, \binits{R.}},
\bauthor{\bsnm{{Schlickeiser}}, \binits{R.}}:
\batitle{{Self-consistent Alfv{\'e}n-wave transmission and test-particle
  acceleration at parallel shocks}}.
\bjtitle{Astron. Astrophys.}
\bvolume{343},
\bfpage{303}--\blpage{311}
(\byear{1999})
\end{barticle}
\endbibitem

\bibitem{Sea2017}
\begin{barticle}
\bauthor{\bsnm{{Sarbadhicary}}, \binits{S.K.}},
\bauthor{\bsnm{{Badenes}}, \binits{C.}},
\bauthor{\bsnm{{Chomiuk}}, \binits{L.}},
\bauthor{\bsnm{{Caprioli}}, \binits{D.}},
\bauthor{\bsnm{{Huizenga}}, \binits{D.}}:
\batitle{{Supernova remnants in the Local Group - I. A model for the radio
  luminosity function and visibility times of supernova remnants}}.
\bjtitle{Mon. Not. R. Astron. Soc.}
\bvolume{464}(\bissue{2}),
\bfpage{2326}--\blpage{2340}
(\byear{2017})
{\href{https://arxiv.org/abs/1605.04923}{{arXiv:1605.04923}}}
{[astro-ph.HE]}.
\doiurl{10.1093/Mon. Not. R. Astron. Soc./stw2566}
\end{barticle}
\endbibitem

\bibitem{Leahy2022}
\begin{barticle}
\bauthor{\bsnm{{Leahy}}, \binits{D.A.}},
\bauthor{\bsnm{{Merrick}}, \binits{F.}},
\bauthor{\bsnm{{Filipovi{\'c}}}, \binits{M.}}:
\batitle{{Radio Emission from Supernova Remnants: Model Comparison with
  Observations}}.
\bjtitle{Universe}
\bvolume{8}(\bissue{12}),
\bfpage{653}
(\byear{2022}).
\doiurl{10.3390/universe8120653}
\end{barticle}
\endbibitem

\bibitem{Pea2018}
\begin{barticle}
\bauthor{\bsnm{{Pavlovi{\'c}}}, \binits{M.Z.}},
\bauthor{\bsnm{{Uro{\v{s}}evi{\'c}}}, \binits{D.}},
\bauthor{\bsnm{{Arbutina}}, \binits{B.}},
\bauthor{\bsnm{{Orlando}}, \binits{S.}},
\bauthor{\bsnm{{Maxted}}, \binits{N.}},
\bauthor{\bsnm{{Filipovi{\'c}}}, \binits{M.D.}}:
\batitle{{Radio Evolution of Supernova Remnants Including Nonlinear Particle
  Acceleration: Insights from Hydrodynamic Simulations}}.
\bjtitle{Astrophys. J.}
\bvolume{852}(\bissue{2}),
\bfpage{84}
(\byear{2018})
{\href{https://arxiv.org/abs/1711.06013}{{arXiv:1711.06013}}}
{[astro-ph.HE]}.
\doiurl{10.3847/1538-4357/aaa1e6}
\end{barticle}
\endbibitem

\bibitem{Gh2007}
\begin{barticle}
\bauthor{\bsnm{{Ghavamian}}, \binits{P.}},
\bauthor{\bsnm{{Laming}}, \binits{J.M.}},
\bauthor{\bsnm{{Rakowski}}, \binits{C.E.}}:
\batitle{{A Physical Relationship between Electron-Proton Temperature
  Equilibration and Mach Number in Fast Collisionless Shocks}}.
\bjtitle{Astrophys. J. Lett.}
\bvolume{654}(\bissue{1}),
\bfpage{69}--\blpage{72}
(\byear{2007})
{\href{https://arxiv.org/abs/astro-ph/0611306}{{arXiv:astro-ph/0611306}}}
{[astro-ph]}.
\doiurl{10.1086/510740}
\end{barticle}
\endbibitem

\bibitem{Gh2013}
\begin{barticle}
\bauthor{\bsnm{{Ghavamian}}, \binits{P.}},
\bauthor{\bsnm{{Schwartz}}, \binits{S.J.}},
\bauthor{\bsnm{{Mitchell}}, \binits{J.}},
\bauthor{\bsnm{{Masters}}, \binits{A.}},
\bauthor{\bsnm{{Laming}}, \binits{J.M.}}:
\batitle{{Electron-Ion Temperature Equilibration in Collisionless Shocks: The
  Supernova Remnant-Solar Wind Connection}}.
\bjtitle{Space Sci. Rev.}
\bvolume{178}(\bissue{2-4}),
\bfpage{633}--\blpage{663}
(\byear{2013})
{\href{https://arxiv.org/abs/1305.6617}{{arXiv:1305.6617}}}
{[astro-ph.GA]}.
\doiurl{10.1007/s11214-013-9999-0}
\end{barticle}
\endbibitem

\bibitem{Gh2016}
\begin{bchapter}
\bauthor{\bsnm{{Ghavamian}}, \binits{P.}}:
\bctitle{{Electron-ion thermal equilibration in collisionless shocks}}.
In: \bbtitle{Supernova Remnants: An Odyssey in Space After Stellar Death},
p. \bfpage{68}
(\byear{2016})
\end{bchapter}
\endbibitem

\bibitem{Ray2023}
\begin{barticle}
\bauthor{\bsnm{{Raymond}}, \binits{J.C.}},
\bauthor{\bsnm{{Ghavamian}}, \binits{P.}},
\bauthor{\bsnm{{Bohdan}}, \binits{A.}},
\bauthor{\bsnm{{Ryu}}, \binits{D.}},
\bauthor{\bsnm{{Niemiec}}, \binits{J.}},
\bauthor{\bsnm{{Sironi}}, \binits{L.}},
\bauthor{\bsnm{{Tran}}, \binits{A.}},
\bauthor{\bsnm{{Amato}}, \binits{E.}},
\bauthor{\bsnm{{Hoshino}}, \binits{M.}},
\bauthor{\bsnm{{Pohl}}, \binits{M.}},
\bauthor{\bsnm{{Amano}}, \binits{T.}},
\bauthor{\bsnm{{Fiuza}}, \binits{F.}}:
\batitle{{Electron-Ion Temperature Ratio in Astrophysical Shocks}}.
\bjtitle{Astrophys. J.}
\bvolume{949}(\bissue{2}),
\bfpage{50}
(\byear{2023})
{\href{https://arxiv.org/abs/2303.08849}{{arXiv:2303.08849}}}
{[astro-ph.GA]}.
\doiurl{10.3847/1538-4357/acc528}
\end{barticle}
\endbibitem

\bibitem{Bell2013}
\begin{barticle}
\bauthor{\bsnm{{Bell}}, \binits{A.R.}},
\bauthor{\bsnm{{Schure}}, \binits{K.M.}},
\bauthor{\bsnm{{Reville}}, \binits{B.}},
\bauthor{\bsnm{{Giacinti}}, \binits{G.}}:
\batitle{{Cosmic-ray acceleration and escape from supernova remnants}}.
\bjtitle{Mon. Not. R. Astron. Soc.}
\bvolume{431}(\bissue{1}),
\bfpage{415}--\blpage{429}
(\byear{2013})
{\href{https://arxiv.org/abs/1301.7264}{{arXiv:1301.7264}}}
{[astro-ph.HE]}.
\doiurl{10.1093/Mon. Not. R. Astron. Soc./stt179}
\end{barticle}
\endbibitem

\bibitem{Caprioli2010}
\begin{barticle}
\bauthor{\bsnm{{Caprioli}}, \binits{D.}},
\bauthor{\bsnm{{Amato}}, \binits{E.}},
\bauthor{\bsnm{{Blasi}}, \binits{P.}}:
\batitle{{Non-linear diffusive shock acceleration with free-escape boundary}}.
\bjtitle{Astroparticle Physics}
\bvolume{33}(\bissue{5-6}),
\bfpage{307}--\blpage{311}
(\byear{2010})
{\href{https://arxiv.org/abs/0912.2714}{{arXiv:0912.2714}}}
{[astro-ph.HE]}.
\doiurl{10.1016/j.astropartphys.2010.03.001}
\end{barticle}
\endbibitem

\bibitem{Blasi2010}
\begin{barticle}
\bauthor{\bsnm{{Blasi}}, \binits{P.}}:
\batitle{{Shock acceleration of electrons in the presence of synchrotron losses
  - I. Test-particle theory}}.
\bjtitle{Mon. Not. R. Astron. Soc.}
\bvolume{402}(\bissue{4}),
\bfpage{2807}--\blpage{2816}
(\byear{2010})
{\href{https://arxiv.org/abs/0912.2053}{{arXiv:0912.2053}}}
{[astro-ph.HE]}.
\doiurl{10.1111/j.1365-2966.2009.16110.x}
\end{barticle}
\endbibitem

\bibitem{ZA2007}
\begin{barticle}
\bauthor{\bsnm{{Zirakashvili}}, \binits{V.N.}},
\bauthor{\bsnm{{Aharonian}}, \binits{F.}}:
\batitle{{Analytical solutions for energy spectra of electrons accelerated by
  nonrelativistic shock-waves in shell type supernova remnants}}.
\bjtitle{Astron. Astrophys.}
\bvolume{465}(\bissue{3}),
\bfpage{695}--\blpage{702}
(\byear{2007})
{\href{https://arxiv.org/abs/astro-ph/0612717}{{arXiv:astro-ph/0612717}}}
{[astro-ph]}.
\doiurl{10.1051/0004-6361:20066494}
\end{barticle}
\endbibitem

\end{thebibliography}


\end{document}